\documentclass[prb,aps,twocolumn,showpacs]{revtex4}
\usepackage[dvips]{graphicx}
\usepackage{psfrag}
\usepackage{amsmath}

\begin{document}

\title{Transport through single-level quantum dot in a magnetic field}

\author{Riccardo Gezzi, Andreas Dirks, Thomas Pruschke}

\affiliation{Institut f\"ur Theoretische Physik, Universit\"at G\"ottingen,
  Friedrich-Hund-Platz~1, D-37077 G\"ottingen, Germany}

\date{\today}

\begin{abstract}
We study the effect of an external magnetic field on the 
transport properties of a quantum dot 
using a recently developed extension of the functional renormalization group
approach to non-equilibrium situations.
We discuss in particular the interplay and competition of the different energy
scales of the dot and the magnetic field on the stationary non-equilibrium
current and conductance.
As rather interesting behavior we find a switching behavior of the magnetic field
for intermediate correlations and bias voltage.
\end{abstract}
\maketitle
\section{Introduction}\label{sec:intro}
The investigation of transport through mesoscopic systems has developed into
a very active research field in condensed matter during the past decade
due to their possible relevance for next-generation electronic devices
and quantum computing.\cite{goldhaber:2003,zutic:2004}
The advance in preparation and nano-structuring of layered semiconductors\cite{kouwenhoven01} 
or the handling of molecules respectively nano-tubes has lead to an increasing
amount of experimental knowledge about such systems.\cite{gershenson:2006} 

The simplest realization of a mesoscopic system is the quantum dot.\cite{kastner92,kouwenhoven97,kouwenhoven01}
It can be viewed as artificial atom coupled to an external bath, whose properties
can be precisely manipulated over a wide range.\cite{kouwenhoven01}
The transport properties of quantum dots in the linear response regime are
very well understood from the experimental as well as theoretical point of view.\cite{kouwenhoven01} 
On the other hand, a reliable theoretical description of even the stationary
transport in non-equilibrium is still a considerable challenge.

In the present work we study the influence of an external magnetic
field $B$ on the stationary transport properties
of a single-level quantum dot subject to a  bias voltage $V_B$
at $T=0$. 
Quantum dots in external magnetic field have been the subject of
interest for some time and experimental studies of these systems
have been performed by several groups.\cite{defranceschi:02,goldhaber:98,ralph:92,schmid:00}
From a theoretical point of view, non-equilibrium properties
in magnetic field were investigated by Meir and Wingreen
\cite{Meir1:92}
combining different methods such as noncrossing approximation (NCA), 
equations of motion (EOM) and variational wavelenght approach 
in the limit in which the Coulomb repulsion $U$ is very large.
Rosch et al.\cite{rosch:05} used a perturbative renormalization
group, which permits the description of the transport properties 
in single-level quantum dots for large bias and magnetic fields. 
K\"onig et al.\cite{koenig:96}
studied tunneling through a single-level quantum dot in the
presence of strong Coulomb repulsion  beyond the perturbative regime
by means of a real-time diagrammatic formulation. 
However, a theory that allows to access intermediate coupling,
bias voltage and magnetic field strenghts on a unique footing
is missing so far.

Here, we use as theoretical approach the functional renormalization group
(fRG)\cite{salmhoferbuch} extended to non-equilibrium,
which we have introduced in a recent publication.\cite{Gezzi:07}
We discuss how the current $J$ and the
differential conductance $G$ are affected by $B$ and by the competition
between magnetic field and bias voltage. We show in particular
that $B$ is responsible for a switching behaviour in $J$ as function of $V_B.$
Interesting are also the individual contributions of spin up and down electrons, 
split by the presence of the magnetic field, to the transport parameters. 
To test our non-equilibrium fRG we have moreover studied, as limiting case,
the equilibrium situation $V_B=0$, in order to compare the results of
the imaginary-time fRG, which has been shown to provide a very good description
of transport through mesoscopic systems in the linear response regime.\cite{meden:06} 

The paper is organized as follows: The next section is dedicated
to a brief description of the model we use, namely the single impurity
Anderson model (SIAM).
Based on the derivation in \textcite{Gezzi:07} we will provide an expression
for the non-equilibrium fRG equations studied in this work. 
In Sec.\ III we present our results for the stationary transport properties.
The limit $V_B\to0$ is investigated first to make contact to previous work and
find the regime, where our method is applicable.
We then discuss how the transport parameters current $J$ and conductance $G$
behave as functions of an applied bias-voltage with and without magnetic field.
Finally we consider the range of applicability of the non-equilibrium 
fRG in the presence of $B$. A summary and conclusions will finish the paper.

\section{Flow equations for single-level quantum dot in magnetic field}
We consider in the following a quantum dot consisting of a single level 
$\varepsilon_{\sigma}$ with spin quantum number $\sigma=\pm1$, coupled to left ($L$) and right ($R$) leads. The electronic
states in the leads are described by a continuum of single-particle levels
with dispersion $\varepsilon_{\vec{k}\sigma\alpha}$, where $\vec{k}$
denotes the wave vector and $\alpha=L$, $R$.
Furthermore, we assume that the leads are always in equilibrium. The dot
and the leads are coupled through an energy and spin independent hybridization
$V_{\alpha}$. Finally, if two electrons occupy the dot, they
experience a Coulomb repulsion $U$. This situation is described by the long-known
single impurity Anderson model, given by
the Hamiltonian\cite{anderson:61}
\begin{eqnarray}
H & = &
\sum_{\vec{k}\sigma\alpha} \varepsilon_{\vec{k}\sigma\alpha}
c^\dag_{\vec{k}\sigma\alpha} c^{\phantom{\dag}}_{\vec{k}\sigma\alpha}
\nonumber\\
&& + \sum_\sigma \varepsilon_{\sigma} d^\dag_\sigma d^{\phantom{\dag}}_\sigma + 
U\left(n_\uparrow-\frac{1}{2}\right)\left(n_\downarrow-\frac{1}{2}\right)
\nonumber\\
\label{eq:siam} 
&& +\frac{1}{\sqrt{N}}
\sum_{\vec{k}\sigma\alpha}\left[V_{\alpha}c^\dag_{\vec{k}\sigma\alpha}d^{\phantom{\dag}}_\sigma+h.c.\right] \;.
\end{eqnarray}
The left and right reservoirs can
have different chemical potentials $\mu_\alpha$ through an applied
bias voltage $V_B=\mu_L-\mu_R$. 
An external magnetic field is taken into account by the Zeeman splitting of
the dot level, 
i.e.\ $\varepsilon_{\sigma}=V_G\pm\sigma B/2$,
where we introduced the gate voltage $V_G$, which controls the filling on the dot.
Since the magnetic fields applied are much smaller than the Fermi energy of the
leads, its effect on the electrons in the leads can be ignored for the present purpose.

The detailed derivation of the fRG flow equations in a general non-equilibrium
situation has been already discussed in Ref.\ \onlinecite{Gezzi:07}. Since one obtains
an infinite hierarchy of differential equations for the irreducible $n$-particle
vertices depending on a cutoff parameter $\Lambda$ in this approach,\cite{Salmhofer:2001,hedden04,Gezzi:07} a truncation is necessary for actual calculations, which we realize by setting the
three-particle vertex $\gamma_3^{\Lambda}\equiv0$. In addition to this truncation of the
hierarchy of differential equations, we further neglect the energy
dependence of the vertex function $\gamma_2^{\Lambda}\equiv\gamma^\Lambda$, which results in an energy
independent single particle self-energy $\Sigma^{\Lambda}$. Note that
in the non-equilibrium approach, all quantities become tensors with respect to
the branches $+$ and $-$ of the Keldysh contour.\cite{Keldysh,Gezzi:07} 
In particular the Green function and self-energy are matrices, which we
denote by $\hat{G}$ and $\hat{\Sigma}$, respectively.

Last but not least, we have to specify how the cutoff $\Lambda$ is introduced.
As usual\cite{hedden04} we choose a $\Theta$-cutoff on the level of the non-interacting dot
here, i.e.\ 
$$
\hat{G}_{d\sigma,0}^{\Lambda}(\omega)=\Theta(\Lambda-|\omega|)\hat{G}_{d\sigma,0}
$$
with $\hat{G}_{d\sigma,0}(\omega)$ the Green function matrix of the dot for $U=0$.
Since in this case we have to deal
with a non-interacting system, the corresponding expressions can be
derived straightforwardly. For simplicity we assume that the dispersions
$\varepsilon_{\vec{k}\sigma\alpha}$ and hybridizations
$V_{\alpha}$ are identical for $\alpha=L$ and $\alpha=R$. In that case the coupling between dot and leads
is characterized by the quantities
\begin{equation}\label{eq:Gamma}
\Gamma_\alpha=\pi|V|^2N_{\rm F}\equiv\frac{\Gamma}{2}\;\;,
\end{equation}
with $N_{\rm F}$ the local density of states of the leads at the dot site.
The result for $\hat{G}_{d\sigma,0}(\omega)$ then reads
\cite{Gezzi:07}:
\begin{eqnarray}\label{eq:kulp1}G_{d\sigma,0}^{--}(\omega) &=&
\frac{\omega-\varepsilon_{\sigma}-i\Gamma\left[1 - 
  f_L(\omega)- f_R(\omega)\right]} {(\omega-\varepsilon_{\sigma})^2+\Gamma^2}\;, \\
\label{eq:kulp2}G_{d\sigma,0}^{++}(\omega)&=&-[G_{d\sigma,0}^{--}(\omega)]^*\;,\\
\label{eq:pulp1}G_{d\sigma,0}^{-+}(\omega)&=&i\frac{\Gamma\left[
  f_L(\omega)+f_R(\omega)\right]}{(\omega-\varepsilon_{\sigma})^2+\Gamma^2}\;,\\
\label{eq:tulp1}G_{d\sigma,0}^{+-}(\omega)&=&-i\frac{\Gamma\left[
  f_L(-\omega)+f_R(-\omega)\right]}{(\omega-\varepsilon_{\sigma})^2+\Gamma^2}\;,
\end{eqnarray}
where $f_\alpha(\pm\omega):=f\left(\pm(\omega-\mu_\alpha)\right)$ 
are the Fermi functions of the leads.

With these definitions and approximations, the resulting system of differential equations for
the spin-dependent single particle self-energy and two-particle vertex we
are going to integrate is given by (for details see e.g.\ App.\ B in Ref.\ \onlinecite{Gezzi:07})
\begin{widetext}
\begin{equation}\label{eq:gamma1f2}
\frac{d}{d\Lambda}\Sigma^{\alpha\beta,\Lambda}_{\sigma}
= -\frac{1}{2\pi}\sum\limits_{\sigma'}
\sum\limits_{\omega=\pm\Lambda}\sum\limits_{\mu\nu} \tilde G^{\mu\nu,\Lambda}_{\sigma'}(\omega)
\gamma^{\alpha\nu\beta\mu,\Lambda}_{\sigma,\sigma',\sigma,\sigma'}\;,
\end{equation}
\begin{eqnarray}
 \frac{d}{d\Lambda} \gamma^{\alpha \beta \gamma \delta, \Lambda}_{\sigma_1',\sigma_2';\sigma_1,\sigma_2}
 = \frac{1}{4\pi}\sum_{\omega=\pm \Lambda}
 \sum_{\sigma_3,\sigma_4}\sum_{\mu,\nu\rho,\eta} 
\bigg(&\!\!\!\tilde G^{\rho \eta,\Lambda}_{\sigma_3}(-\omega) \tilde G^{\nu \mu,\Lambda}_{\sigma_4}(\omega)&\!\!\!
\gamma^{\alpha \beta \rho \nu,\Lambda}_{\sigma_1',\sigma_2';\sigma_3,\sigma_4}  
\gamma^{\eta \mu \gamma\delta,\Lambda}_{\sigma_3,\sigma_4;\sigma_1,\sigma_2}
- \nonumber\\
&\!\!\! \tilde G^{\eta \rho,\Lambda}_{\sigma_3}(\omega) \tilde G^{\nu \mu,\Lambda}_{\sigma_4}(\omega)\Big[ &\!\!\!
\gamma^{\alpha \mu \gamma \eta,\Lambda}_{\sigma_1',\sigma_4;\sigma_1,\sigma_3}  
\gamma^{\rho \beta \nu \delta,\Lambda}_{\sigma_3,\sigma_2';\sigma_4,\sigma_2} 
+
\gamma^{\alpha \rho \gamma \nu,\Lambda}_{\sigma_1',\sigma_3;\sigma_1,\sigma_4}  
\gamma^{\mu \beta \eta \delta,\Lambda}_{\sigma_4,\sigma_2';\sigma_3,\sigma_2} -\nonumber\\
\label{eqq2}
&\!\!\!&\!\!\!
\gamma^{\beta \mu \gamma \eta,\Lambda}_{\sigma_2',\sigma_4;\sigma_1,\sigma_3}\gamma^{\rho \alpha \nu
  \delta,\Lambda}_{\sigma_3,\sigma_1';\sigma_4,\sigma_2}-
\gamma^{\beta \rho \gamma \nu,\Lambda}_{\sigma_2',\sigma_3;\sigma_1,\sigma_4}\gamma^{\mu \alpha \eta
  \delta,\Lambda}_{\sigma_4,\sigma_1';\sigma_3,\sigma_2}
\Big] \bigg)\;\;.
\end{eqnarray}
\end{widetext}
In expressions (\ref{eq:gamma1f2}) and (\ref{eqq2})
$$
\tilde{G}_\sigma^{\mu\nu,\Lambda}(\omega)=\left[\frac{1}{\hat{G}_{d\sigma,0}(\omega)^{-1}-\hat{\Sigma}_\sigma^{\Lambda}}\right]^{\mu\nu}\;\;,
$$
and the upper small greek indices refer to the branches of the Keldysh contour.
The initial conditions at $\Lambda=\infty$ are are given by
$$
\Sigma_{\sigma}^{\Lambda=\infty}=0
$$
and
$$
\gamma_{\sigma_1',\sigma_2';\sigma_1,\sigma_2}^{\alpha\alpha\alpha\alpha,\Lambda=\infty}=
i\alpha U (\delta_{\sigma_1,\sigma_1'}\delta_{\sigma_2,\sigma_2'}-
\delta_{\sigma_1,\sigma_2'}\delta_{\sigma_2,\sigma_1'})\;\;.
$$
All other components of $\gamma^{\Lambda=\infty}=0$. The integration of the
equations  (\ref{eq:gamma1f2}) and (\ref{eqq2}) has to be done until
$\Lambda=0$ is reached.

Compared to the system obtained in Ref.\ \onlinecite{Gezzi:07}
without external magnetic field the set of eqs.\ (\ref{eq:gamma1f2}) and (\ref{eqq2}) 
show a more complicated structure which manifests itself in a
spin-dependent flow for the selfenergy and the vertex. 
As $V_B \to 0,$ it can be shown \cite{meden:06} that the
vertex $\gamma_2$ can be parametrized as
$$\gamma^{\alpha \beta \gamma
\delta,\Lambda}_{\sigma_1',\sigma_2';\sigma_1,\sigma_2}=\delta_{\sigma_1',\sigma_1} 
\delta_{\sigma_2',\sigma_2} U^{\alpha \beta \gamma \delta,\Lambda}-
\delta_{\sigma_2',\sigma_1}\delta_{\sigma_1',\sigma_2} U^{\beta \alpha
  \gamma \delta,\Lambda},$$
with spin-independent interaction parameters $U$.
It is thus tempting to use the same parametrization out of equilibrium, too,
resulting in the simpler set of equations
\begin{widetext}
\begin{equation}
  \label{eq:sigme_full}
  \frac{d}{d\Lambda}\Sigma^{\alpha\beta,\Lambda}_{\sigma}=
-\frac{1}{2\pi}\sum\limits_{\omega=\pm\Lambda,\gamma,\delta}\left[\left(\tilde G_{\sigma}^{\gamma\delta,\Lambda}(\omega)+
\tilde G_{-\sigma}^{\gamma\delta,\Lambda}(\omega)\right)U^{\alpha\beta\gamma\delta,\Lambda}
-\tilde G_{\sigma}^{\gamma\delta,\Lambda}(\omega)U^{\beta\alpha\gamma\delta,\Lambda}
\right]\;\;,
\end{equation}
\begin{eqnarray}
\frac{d}{d\Lambda} U^{\alpha \beta \gamma \delta,\Lambda}
 = \frac{1}{4\pi}\sum_{\omega=\pm\Lambda}\sum_{\mu,\nu\rho,\eta} 
\biggr(
\tilde G^{\rho \eta,\Lambda}_{-\sigma}(-\omega) 
\tilde G^{\nu\mu,\Lambda}_{\sigma}(\omega) 
U^{\alpha \beta \eta \nu,\Lambda} U^{\rho \mu \gamma\delta,\Lambda}
+ 
\tilde G^{\rho \eta,\Lambda}_{\sigma}(-\omega) 
\tilde G^{\nu \mu,\Lambda}_{-\sigma}(\omega)
U^{\beta \alpha \eta \nu,\Lambda} U^{\mu\rho\gamma\delta,\Lambda}-\nonumber\\
\hspace*{-1cm}\left[
\tilde G^{\eta \rho,\Lambda}_{\sigma}(\omega)
\tilde G^{\nu\mu,\Lambda}_{\sigma}(\omega)
+
\tilde G^{\eta \rho,\Lambda}_{-\sigma}(\omega) 
\tilde G^{\nu\mu,\Lambda}_{-\sigma}(\omega)\right] 
U^{\alpha \mu \gamma \eta,\Lambda} U^{\rho \beta \nu \delta,\Lambda}- 
\tilde G^{\eta \rho,\Lambda}_{\sigma}(\omega) 
\tilde G^{\nu\mu,\Lambda}_{\sigma}(\omega)
U^{\alpha \mu \gamma \eta,\Lambda} U^{\beta \rho \nu \delta,\Lambda}-\nonumber\\
\tilde G^{\eta \rho,\Lambda}_{-\sigma}(\omega) 
\tilde G^{\nu\mu,\Lambda}_{-\sigma}(\omega)
U^{\mu \alpha \gamma \eta,\Lambda} U^{\rho \beta \nu \delta,\Lambda}+
\left[
\tilde G^{\eta \rho,\Lambda}_{\sigma}(\omega)
\tilde G^{\nu\mu,\Lambda}_{\sigma}(\omega)+
\tilde G^{\eta \rho,\Lambda}_{-\sigma}(\omega)
\tilde G^{\nu\mu,\Lambda}_{-\sigma}(\omega)
\right]
U^{\alpha \rho \gamma \nu,\Lambda} U^{\mu \beta \eta \delta,\Lambda}-\nonumber\\
\tilde G^{\eta \rho,\Lambda}_{\sigma}(\omega)
\tilde G^{\nu\mu,\Lambda}_{\sigma}(\omega)
U^{\alpha \rho \gamma \nu,\Lambda} U^{\beta \mu \eta \delta,\Lambda}-
\tilde G^{\eta \rho,\Lambda}_{-\sigma}(\omega)
\tilde G^{\nu\mu,\Lambda}_{-\sigma}(\omega)
U^{\rho \alpha \gamma \nu,\Lambda} U^{\mu \beta \eta \delta,\Lambda}-\nonumber\\
\tilde G^{\eta \rho,\Lambda}_{\sigma}(\omega)
\tilde G^{\nu \mu,\Lambda}_{-\sigma}(\omega)
U^{\mu \beta \gamma \eta,\Lambda}U^{\alpha \rho \nu \delta,\Lambda}+
\tilde G^{\eta \rho,\Lambda}_{-\sigma}(\omega)
\tilde G^{\nu \mu,\Lambda}_{\sigma}(\omega)
U^{\rho \beta \gamma \nu,\Lambda}U^{\alpha \mu \eta \delta,\Lambda}
 \bigg)\;\;.~
\label{eqq3}\end{eqnarray}
\end{widetext}
By investigating the structure of the vertex for the full spin-dependent 
flow at $\Lambda=0$ it turns out, however, that the parametrized version
breaks exact symmetries 
for exchanging indices.
Due to the significantly
reduced numerical effort for integrating Eqs.\ (\ref{eq:sigme_full}) and
(\ref{eqq3}), it remains interesting to examine its value as an approximation.

After integration of system (\ref{eq:gamma1f2}), (\ref{eqq2}) 
or (\ref{eq:sigme_full}), (\ref{eqq3}) we insert the
resulting selfenergy into the Meir-Wingreen formula for the current\cite{Meir:1992, Gezzi:07}
\begin{eqnarray} \label{current}
  J_{\sigma}=\frac{ie\Gamma}{2\pi\hbar}\int
  d\epsilon\left[f_L(\epsilon)-f_R(\epsilon)\right]
  \left(\tilde G^{+-}_{d\sigma}(\epsilon)-
 \tilde G^{-+}_{d\sigma}(\epsilon)\right),
\end{eqnarray}
where $\hat G_{d\sigma}$ denotes the full spin dependent one-particle impurity Green function.

The conductance finally is obtained from (\ref{current}) by numerical differentiation
with respect to $V_B$.

\section{Results}
Together with the initial conditions for self-energy and two-particle vertex
we can now integrate the differential equations using a standard Runge-Kutta
solver.
Although the systems of equations are valid for $T>0$, too, we restrict
our discussion to the case $T=0$ for simplicity, noting that preliminary
results for $T>0$ in the weak-coupling regime are in agreement with previous
studies and, in the linear response regime, NRG results.

\subsection{Test of the parametrized flow}
\begin{figure}[htbp]
\begin{center}
\includegraphics[width=0.45\textwidth,clip]{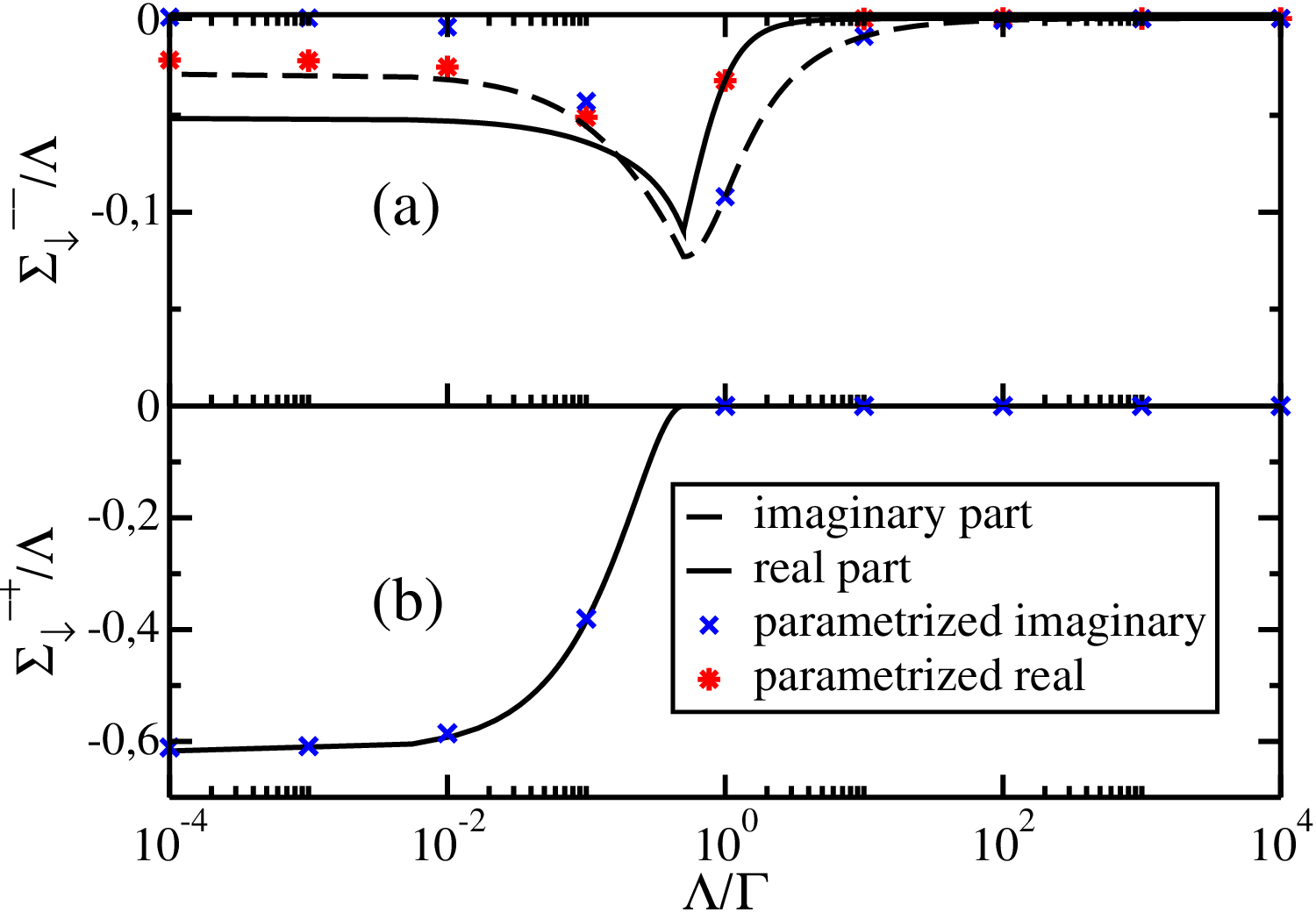}
\caption{(color online) (a) Flow of real (full curve and stars) and
imaginary part (dashed curve and crosses) of 
$\Sigma^{--,\Lambda}_{\downarrow}/\Gamma,$
for $V_G/\Gamma=0,$ $U/\Gamma=5,$ $V_B/\Gamma=1,$ $B/\Gamma=0.116$. 
The continuous lines represent the solutions of
Eqs.\ (\ref{eq:gamma1f2}) and (\ref{eqq2}),
the symbols the parametrized version
Eqs.\ (\ref{eq:sigme_full}) and (\ref{eqq3}).
(b) Flow of $\Sigma^{-+,\Lambda}_{\downarrow}/\Gamma$.\label{fig:A}}
\end{center}
\end{figure}

\begin{figure}[htbp]
\begin{center}
\includegraphics[width=0.4\textwidth,clip]{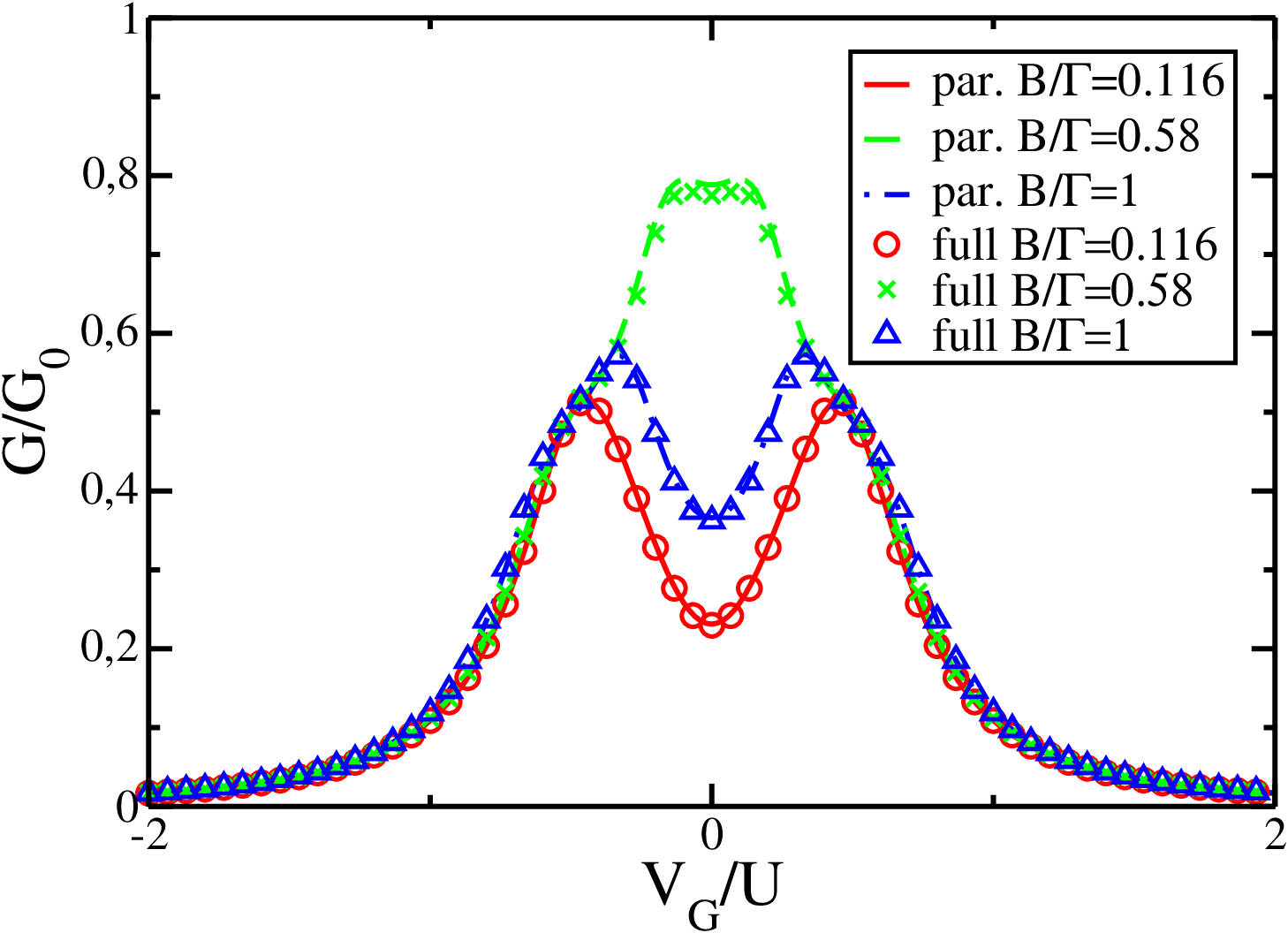}
\caption{(color online) Conductance $G$ normalized 
to $G_0=2e^2/h$ as function of $V_G$ for 
$U/\Gamma=5,$ $V_B/\Gamma=1$ and several values of $B$.\label{fig:B}}
\end{center}
\end{figure}

\begin{figure}[htbp]
\begin{center}
\includegraphics[width=0.4\textwidth,clip]{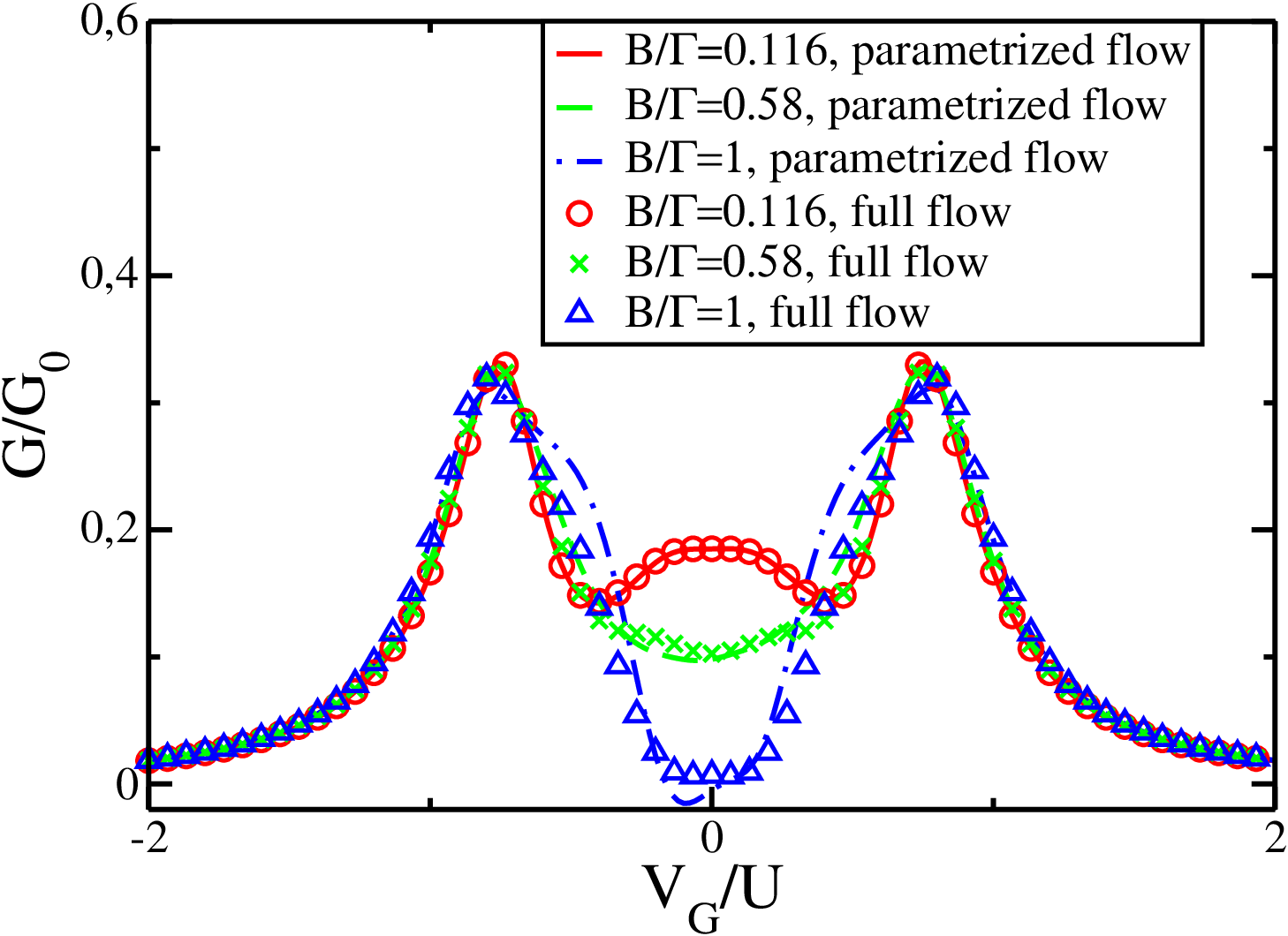}
\caption{(color online) The same parameters as in Fig.\ \ref{fig:B} except
  for $V_B/\Gamma=3$.\label{fig:D}}
\end{center}
\end{figure}

Let us start by
comparing the full flow according to Eqs.\ (\ref{eq:gamma1f2}) and (\ref{eqq2}) 
with the parametrized ones Eqs.\ (\ref{eq:sigme_full}) and (\ref{eqq3}).
We observe a good agreement at small $V_B$ for $\Sigma^{-+,\Lambda}$ (see Fig.\ \ref{fig:A}b),
while deviations appear in $\Sigma^{--,\Lambda}$ (see Fig.\ \ref{fig:A}a). 
However, since 
$|\Sigma^{-+,\Lambda}|\gg|\Sigma^{--,\Lambda}|$, this does not affect
the behaviour of experimentally relevant quantities even for larger $B$,
as can be seen for the case of the conductance $G$ as function of gate voltage
in Fig.\ \ref{fig:B}.
Increasing the bias voltage, the agreement is still good for small $B$ 
(see Fig.\ \ref{fig:D}, full
curve and circles). 
As soon as we increase the magnetic
field, $\mathrm{Im\,}\Sigma^{-+,\Lambda}$, too, shows deviations between the
\begin{figure}[htbp]
\begin{center}
\includegraphics[width=0.45\textwidth,clip]{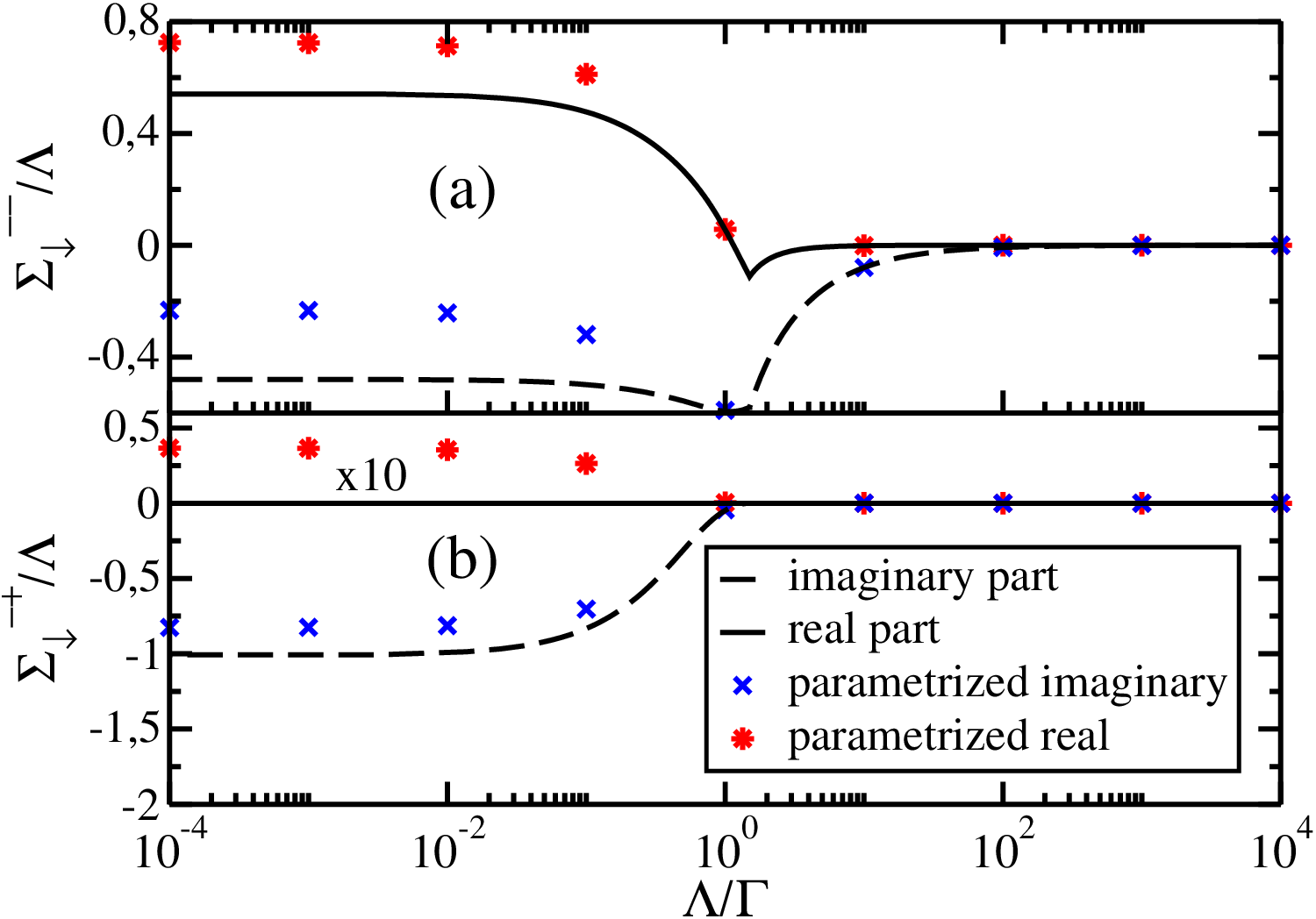}
\caption{(color online) The same parameters as in Fig.\ \ref{fig:A} except for $V_B/\Gamma=3$ and $B/\Gamma=1.$
In (b) the real part of $\Sigma^{-+}$ has been rescaled by a factor ten to make it visible.\label{fig:C}}
\end{center}
\end{figure}
full and parametrized flow (see Fig.\ \ref{fig:C}).
In particular, the flow of the parametrized system 
gives rise to a small real component of $\Sigma^{+-,\Lambda}$ 
at $\Lambda=0$ (Fig.\ \ref{fig:C}b), which actually should not exist,
and leads to unphysical breaking of the particle-hole symmetry in
physical quantities like the conductance, as is visible from the dashed
and dot-dashed curves in Fig.\ \ref{fig:D}.

We thus can conclude that the parametrized set of flow equations is a reasonable
approximation at least for calculating the conductance as long as $V_B$ and
the magnetic field $B$ become not too large. 

\subsection{Equilibrium case}

Before studying the influence of a magnetic field on stationary non-equilibrium
transport at $T=0$, we first present results
for the linear response regime, i.e.\ the limit $V_B=\mu_L-\mu_R \to 0$.
This limit can serve as a test for the non-equilibrium fRG,
because it should reproduce the imaginary-time fRG results.\cite{meden:06}

\begin{figure}[htbp]
\begin{center}
\includegraphics[width=0.45\textwidth,clip]{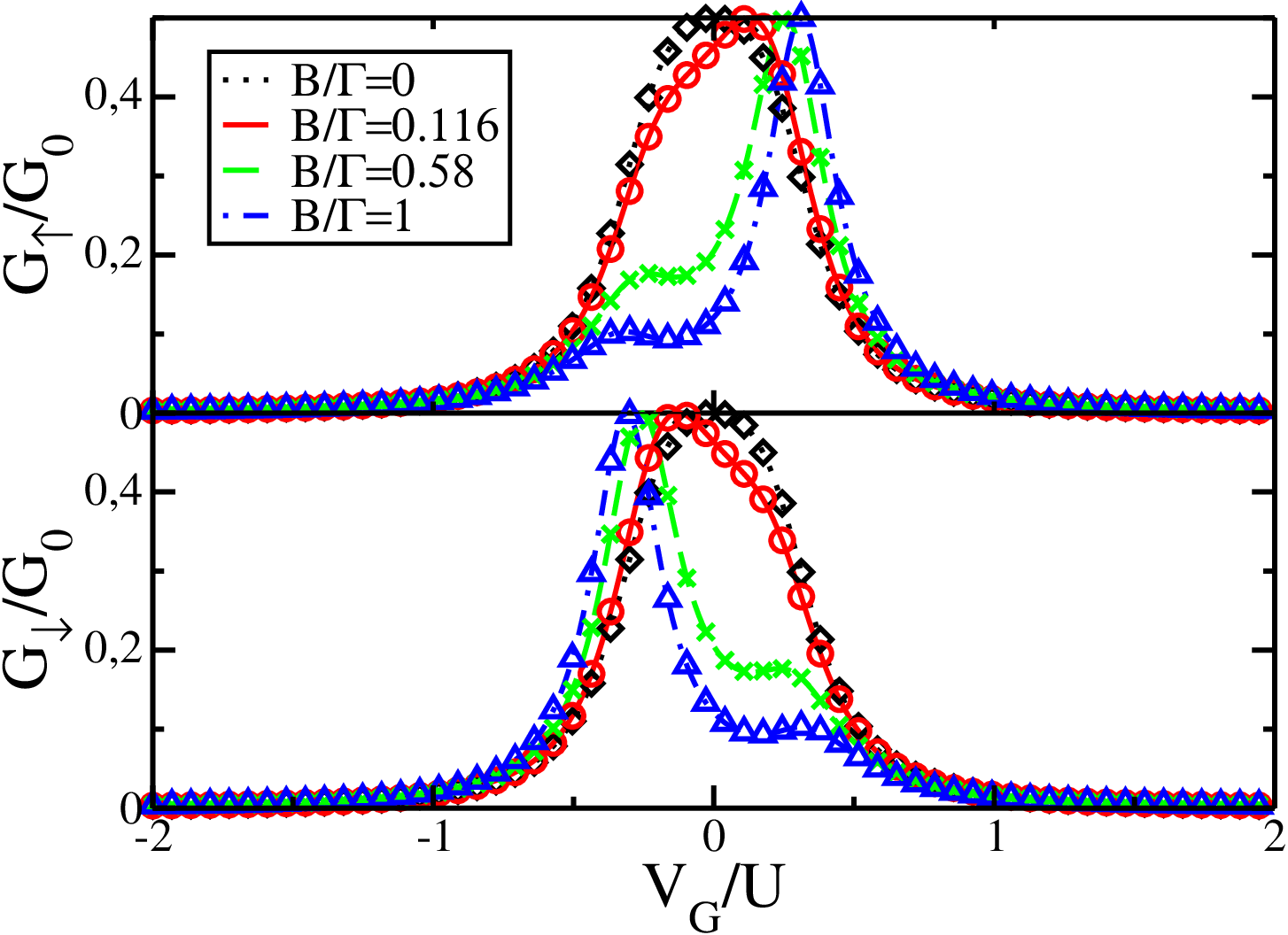}
\caption{(color online) Conductance $G_\sigma$ from the individual spin channels in the limit 
$V_B\to0$ normalized to $G_0=2e^2/h$ as function of $V_G/U$ for 
$U/\Gamma=5$ and several values of $B$. Symbols were obtained from the
non-equilibrium fRG, lines from the imaginary-time fRG of \textcite{meden:06}.\label{fig:F}}
\end{center}
\end{figure}

In Fig.\ \ref{fig:F} we compare 
the contributions $G_\sigma$ to the conductance 
from the individual spin channels calculated with the non-equilibrium
fRG (symbols) as $V_B\to0$ for different values
of $B$ to the results of the imaginary-time fRG obtained by \textcite{meden:06} (lines).
We note, that for $V_B/\Gamma=0$ our calculations perfectly reproduce
the results of \textcite{meden:06}.
We see that, as soon as the magnetic field increases, $G_\sigma$ starts to
split into two peaks
which reflects the field dependent shifts of the
spectral functions for up and down spin. 

We would like to mention here that for too large $U$ 
and magnetic field a crossing of solutions of the differential equations appears, 
and the numerical procedure picks the one on the wrong Riemannian sheet. At present
we do not know how to avoid this numerical instabilty, which limits the
applicability of the method to intermediate coupling parameters. We will
come back to this problem later.
    
For an extensive discussion of the linear response results with an applied magnetic
field we refer the reader to the seminal paper by \textcite{meden:06}.

\subsection{Non-equilibrium}

\begin{figure}[htbp]
\begin{center}
\includegraphics[width=0.45\textwidth,clip]{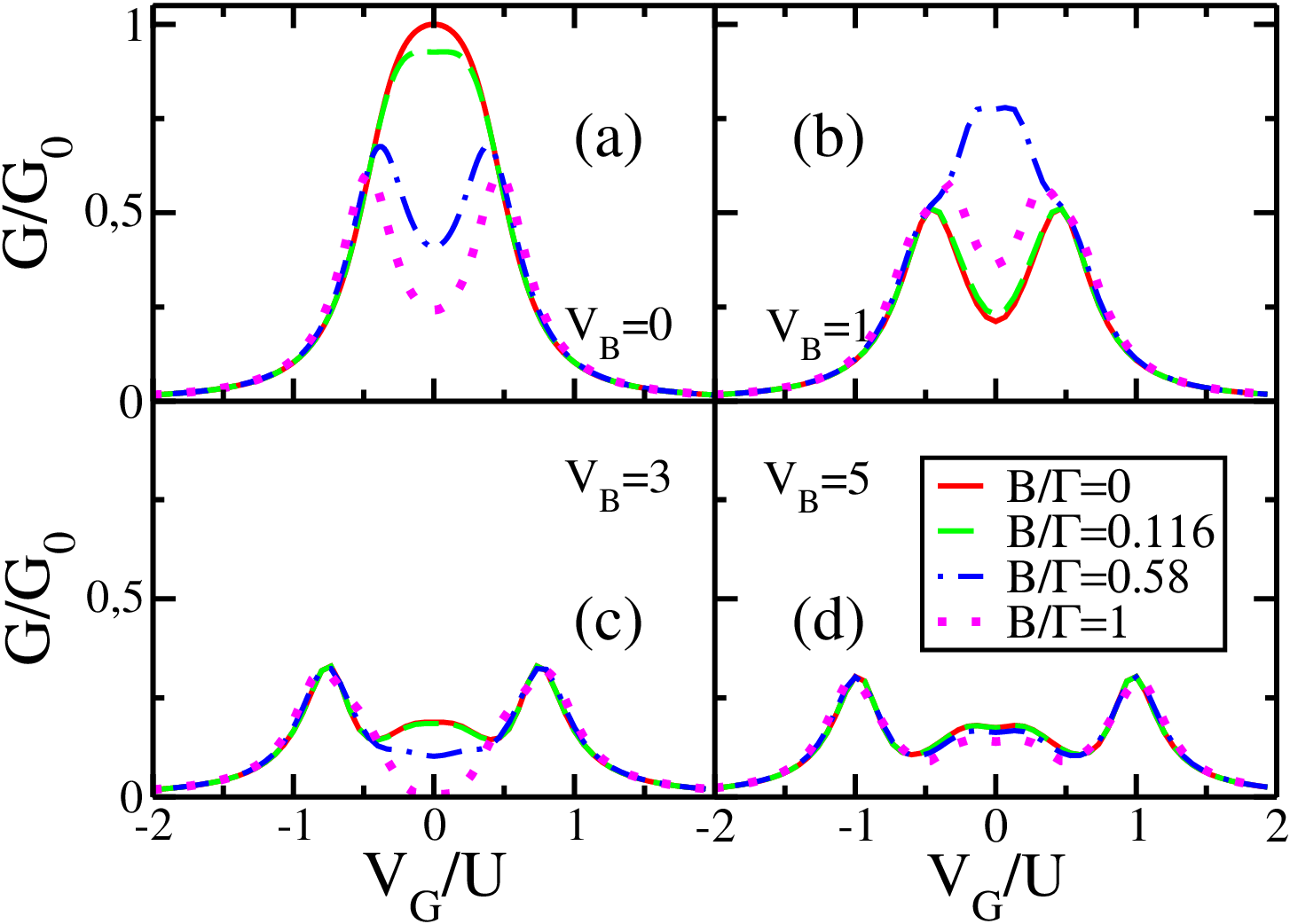}
\caption{(color online) Total conductance $G$ normalized 
to $G_0=2e^2/h$ as function of $V_G/U$ for 
$U/\Gamma=5$, $V_B/\Gamma=0$, $1$, $3$, $5$ and several values of 
$B/\Gamma$.\label{fig:E}}
\end{center}
\end{figure}

Switching on the bias voltage $V_B$ we observe  different behaviours
of the conductance $G$ as a function of the gate voltage $V_G$
depending on the competition between the voltage and 
the magnetic field. In Fig.\ \ref{fig:E}a-d we present $G=G_\uparrow+G_\downarrow$ as
function of $V_G$ for fixed Coulomb interaction $U/\Gamma=5$ and 
$V_B/\Gamma=0$, $1$, $3$, $5$ for different values of $B$.
We observe a drastic change of the conductance due to the interplay of
$V_B $ and $B$. 
In particular, we see that for small bias ($V_B/\Gamma=1$, Fig.\ \ref{fig:E}b)
and fields
$B/\Gamma = 0.116$, $G$ consists of just two peaks separated by $U$. If we
now increase the value of the magnetic field, 
the conductance at small $V_G$ initially strongly increases 
(dashed and dot-dashed curves in Fig.\ \ref{fig:E}b), the peaks disappear 
and a small plateau appears. Increasing $B$ further,
the plateau disappears and we get back the two peaks separated by a rather
deep valley and two shoulders whose spacing is $\Delta\approx U$.

This behaviour can be explained as follows: 
The spectral density for each spin channel is split by $V_B$ into two peaks
moving the spectral weight to higher frequencies and 
decreasing it in the region $V_G\approx 0$.\cite{Gezzi:07} 
Switching on $B$ will lead to a splitting $\sim B$ of each peak, 
due to field induced shifts of the spectral functions from the individual spin contributions.
Thus, increasing $B$, the two outer peaks move
away from each other while the inner ones get
closer and even merge, enhancing the conductance at $V_G\approx 0$. 
As direct consequence we observe
a small plateau for $V_G\approx 0$ (see Fig.\ \ref{fig:E}b). Further increasing
$B$, these contributions will again drift apart, 
leading to a collapse of the conductance and disappearance of the plateau.

Completely different is the behaviour for $V_B/\Gamma=3$ and 
$V_B/\Gamma=5$ (Fig.\ \ref{fig:E}c and d).  
We find a monotonic decrease of G with the magnetic field.
In addition the field dependence is initially weaker than in Fig.\ \ref{fig:E}b.
We interpret this behaviour in the following way: For large $V_B$,
the Fermi window in e.g.\ Eq.\ (\ref{current}) will lead to an averaging
over a large energy region. Thus structures due to the magnetic field 
at too small energies will be washed out, i.e.\ the subtle interplay between
the rearrangement of spectral weight due to $V_B$ and the shifts induced by
the magnetic field $B$ cannot be resolved any more.

\subsection{Current and Conductance as function of the applied bias}

\begin{figure}[htbp]
\begin{center}
\includegraphics[width=0.4\textwidth,clip]{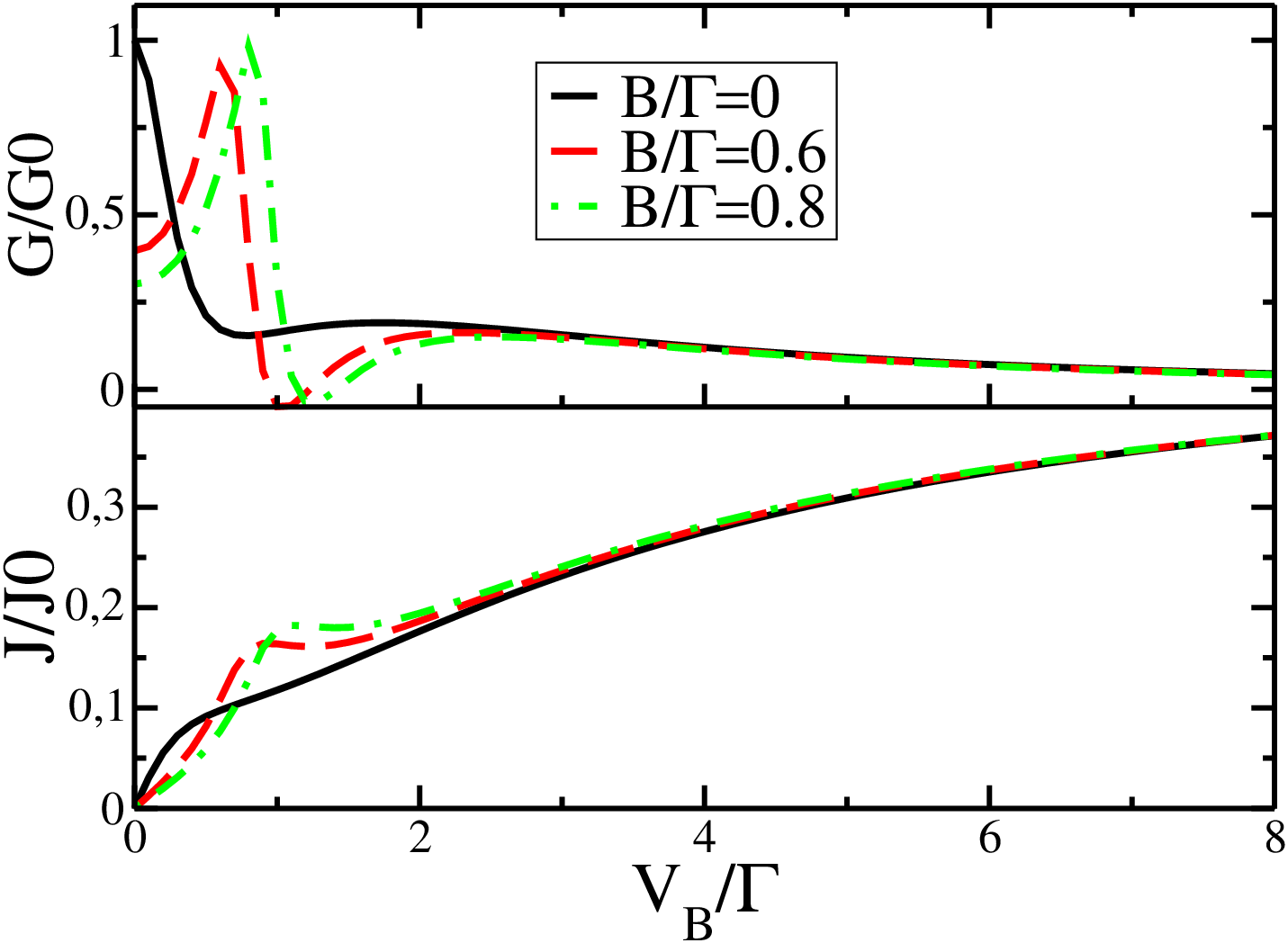}
\end{center}
\caption{(color online) Total current normalized to $J_0=G_0\frac{\Gamma}{e}$ and 
conductance as function of the bias voltage $V_B/\Gamma$
for $V_G/\Gamma=0$, $U/\Gamma=5$ and several values of $B$.\label{fig:G}}
\end{figure}

The non-monotonic behaviour of $G(V_G=0)$ for moderate $V_B$ as function of the
external magnetic field $B$ is an interesting feature
we want to explore in somewhat more detail in the following.
To this end we calculated the current $J$ and the
differential conductance $G=dJ/dV_B,$ at $V_G/\Gamma=0$ and $U/\Gamma=5$
as function of the applied bias
for different values of $B$. The results are collected in Fig.\ \ref{fig:G}.
Compared to the case $B=0$ (full line) we see that a finite
magnetic field $B$ basically induces two features. First, as is
also the case for $B=0$ at larger $U$, we observe a shoulder in the current,
resulting in a small region of
negative differential conductance. More interesting, however, is the appearance
of an almost unitary conductance peak at $V_B\approx B$ for
intermediate magnetic fields. For large fields the features is suppressed
again. 
This behaviour can be explained by noting that when $V_B\approx B$,
the electrochemical potentials $\mu_{L,R}$ of the leads are close to the
split dot levels, respectively, therefore the tunnel probability from
the leads to the dot is enhanced.
For $V_B \gg B$ the curves converge again to the same values, i.e.\ $B$ does
not influence the current any longer, the behavior is dominated by the
now rather wide Fermi window.

Thus we observe that $B$ can act
as a switch, suppressing the current for small $V_B$ and leading to a
steep increase of the current in the voltage range
$V_B \approx B$. 

\subsection{Range of applicability of the fRG}
\begin{figure}[htbp]
\begin{center}
\includegraphics[width=0.45\textwidth,clip]{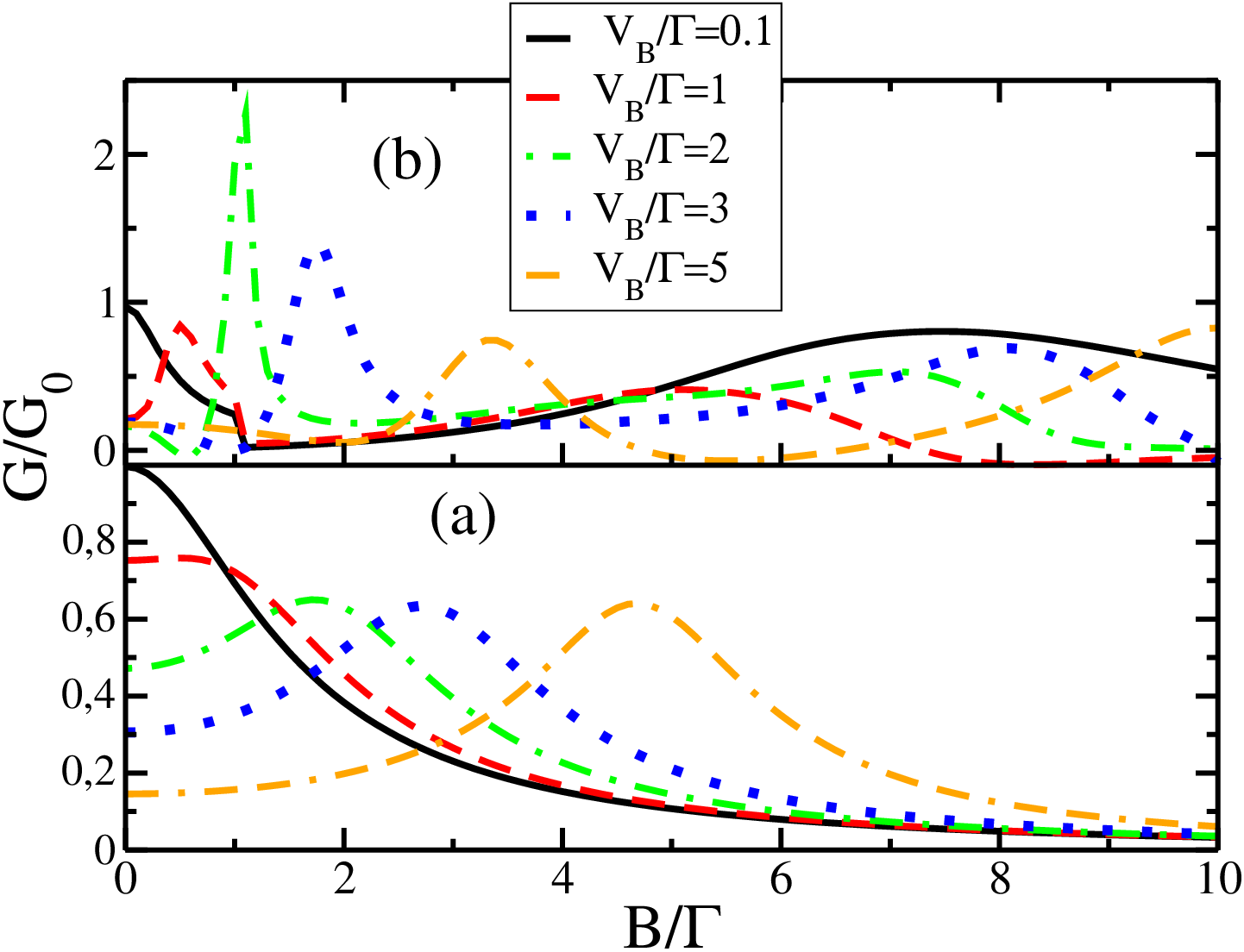}
\end{center}
\caption{(color online) Total conductance normalized to $G_0=2e^2/h$ plotted versus
$B/\Gamma$ for $V_G/\Gamma=0$ and several values of $V_B$. (a)
weak coupling regime $U/\Gamma=1$. (b) strong coupling regime $U/\Gamma=5$.\label{fig:H}}
\end{figure}

Let us finally discuss  in which range of parameters the
non-equilibrium fRG furnishes reliable results.
In Fig.\ \ref{fig:H} we show the transport
parameters plotted as function of the magnetic field for different bias
voltages in the weak ($\frac{U}{\pi\Gamma}<1$, Fig.\ \ref{fig:H}a) and intermediate
coupling regime ($\frac{U}{\pi\Gamma}\gtrsim1$, Fig.\ \ref{fig:H}b).
While for $\frac{U}{\pi\Gamma}<1$ the fRG always converged smoothly,
we see that for $\frac{U}{\pi\Gamma}>1 $ the curves show a discontinuity in $G$
(see Fig.\ \ref{fig:H}8b for $V_B/\Gamma=1$).
Increasing $V_B$ further this discontinuity disappears but for
$V_B/\Gamma=2..3$ the conductance now
overshoots the unitary limit in the range $1<B/\Gamma<2$.

While these results are at first sight very disturbing, they can be traced back
to a crossing of solutions of the differential equations in this parameter regime.
As already mentioned earlier and discussed in detail in Ref.\ \onlinecite{Gezzi:07},
the Runge-Kutta algorithm picks the wrong solution, leading to the observed
pathologies in the physical properties.
At very large bias $V_B \ge U$ the
solution of the differential equations does not show this problem any longer.

This means that our approach is \emph{due to numerical reasons} not reliable in
this parameter range and possibly for larger magnetic fields. We are presently
not aware of any algorithm that can avoid this numerical difficulty related
to the appearance of a pole in the physical Riemannian sheet.
\section{Summary and outlook}
In this work we applied the non-equilibrium fRG approach to the stationary transport
through a single-level quantum dot subject to a constant bias $V_B$ and
magnetic field $B$ at $T=0$.
Besides the truncation of the infinite hierarchy of the fRG differential equations
at the two-particle level we also 
neglected the energy dependence
of the two-particle vertex function $\gamma_2$, resulting in an energy independent
single-particle self energy. Guided by the structure of the fRG equations
in equilibrium, we studied as further possible approximation the neglect
of the spin-dependence in the vertex function $\gamma_2$. This latter
approximation further significantly reduces the complexity of the 
system of differential equations.
We compared the results of this approximation with those from the equations
maintaining the full spin dependence, finding that for not too large $V_B$ and
field $B$ it actually can be used to accurately calculate transport properties.

Although the truncation and especially the neglect of the energy dependence must
be viewed as rather crude approximations, we have shown that one can obtain
reasonable results for the transport parameters $J$ and $G$ up to the
intermediate coupling regime for an extended regime of the parameters $V_B$
and $B$. 
A particularly interesting observation is the switching
behaviour found in the current for intermediate values of $V_B$ and magnetic
field $B$, which we could explain with the interplay of
the different structures we expect in the single-particle spectra as function
of the gate voltage and $B$. 
We also showed that for bias voltage $V_B\to 0$ we
recover, as expected, the linear response results by \textcite{meden:06}.

As an important future step we view the introduction of the energy dependence
in the flow equations of the non-equilibrium fRG. Besides curing certain
deficiencies not discussed here (see \textcite{Gezzi:07}), we expect
that this step will extend the range of applicability of our formalism to
larger magnetic fields and Coulomb interaction. Furthermore, coping with the
full energy dependence of the vertex will eventually enable us to study
time-dependent phenomena off equilibrium, which will open a wide and interesting
field of physical phenomena and applications.

\begin{acknowledgments}
We acknowledge useful conversations with 
K.~Sch\"onhammer,
V.~Meden,
F.~Anders,
J.~Kroha, 
H.~Monien
and
M.~Jarrell.

This work was supported by the DFG through the collaborative research
center SFB 602. Computer resources were provided through the Gesellschaft
f\"ur wissenschaftliche Datenverarbeitung in G\"ottingen and
the Norddeutsche Verbund f\"ur Hoch- und H\"ochstleistungsrechnen.

\end{acknowledgments}

\end{document}